\documentclass[journal=jpcl,manuscript=article]{achemso}
\usepackage{chemformula} % Formula subscripts using \ch{}
\usepackage[T1]{fontenc} % Use modern font encodings

\author{Tarun Gera}
\email{tarungera@iisc.ac.in}
\author{K. L. Sebastian}
\email{kls@iisc.ac.in}

\phone{+91 86306 46746; +91 9632928096}

\affiliation[Indian Institute of Science]
{Department of Inorganic and Physical 
Chemistry, Indian Institute of Science, Bangalore 560012, India}
\title[An \textsf{achemso} demo]
  {A demonstration of the \textsf{achemso} \LaTeX\
   class\footnote{A footnote for the title}}
\abbreviations{IR,NMR,UV}

\usepackage{chemformula} % Formula subscripts using \ch{}
\usepackage[T1]{fontenc} % Use modern font encodings

\title[An \textsf{achemso} demo]
  {Exact Results for the Tavis-Cummings and H$\boldsymbol{\ddot{u}}$ckel Hamiltonians
with Diagonal Disorder\footnote{Disordered Tavis-Cummings Model}}
 
\abbreviations{IR,NMR,UV}
\usepackage[latin9]{inputenc}
\usepackage[english]{babel}
\usepackage{amsmath}
\usepackage{amssymb}
\usepackage{graphicx}
\PassOptionsToPackage{version=3}{mhchem}
\usepackage{mhchem}
\usepackage{babel}
\usepackage{color}
\usepackage{epstopdf}
\usepackage{chemfig}

\usepackage{soul}
\usepackage{amsfonts}
\usepackage{mathrsfs}
\begin{document}
\begin{abstract}{We present an exact method to calculate the electronic states of one electron Hamiltonians with diagonal disorder.  We show that the disorder averaged  one particle Green's function can be calculated directly, using a deterministic complex (non-Hermitian) Hamiltonian.  For this, we assume that the molecular states have a Cauchy (Lorentz) distribution and use the supersymmetric method which has already been used in problems of solid state physics.  Using the method we  find exact solutions to the states of $N$  molecules, confined to a microcavity, for any value of $N$. Our analysis shows that the width of the polaritonic states as a function of $N$ depend on the nature of disorder, and hence can be used to probe the way molecular energy levels are distributed.  We also show how one can find exact results for H$\ddot{u}$ckel type Hamiltonians with on-site, Cauchy disorder and demonstrate its use.}
\end{abstract}

\maketitle
%\section{Introduction}
\newpage
Confining molecules  to micro-cavities of size such that a molecular excitation (electronic or vibrational) is resonant with an electromagnetic mode of the cavity leads to the formation of hybrid light-matter states known as polaritons.   Such confinement 
 can cause modifications of the rates of chemical processes. Rather than giving an extensive list of   the many interesting developments in the area, we refer the reader to the summaries that may be found in Ebbesen et. al\cite{EbbesenPerspective2021}, Hartland and Scholes\cite{Virtual-Issue-JPCL} and Herrera and Owrutsky\cite{Herrera2020}.  
 
Light-matter coupling leads to the formation a  lower and an upper polaritonic
states and the energy difference between the two is the Rabi splitting $\Omega.$ The polaritonic states are an example of collective
phenomena where states of a large number $N$ of molecules combine
to form one state, which then combines with the cavity mode to give
the two polaritonic states. The value of $N$ is very large - roughly
speaking, all the molecules  which are oriented in the direction
of the electric field of the cavity, located at points where the electric
field of the cavity mode is not very small couple. It is the coupling
of this large number of molecules that leads to the large Rabi splitting
whose frequency is a sizeable fraction of the energy of the molecular excitation
involved. The formation
of the polaritonic states is usually understood using the Tavis-Cummings \cite{Tavis1968}
model which has $N$ excitations localized on each molecule, all
having the energy $\epsilon_{a}$ and the cavity mode of energy $\epsilon_{c}$.
In this simple model the $N-1$ modes that remain are completely decoupled
from the radiation and hence are referred to as the dark states.  
While this is the ideal situation, there would be disorder in
the molecular energy levels due to various reasons \cite{Scholes2020}.
The causes for this may be inhomogeneities in the material, different
levels of aggregation of the molecules, solvent fluctuations or dispersal
by solvent. Hence a generalized version of the Tavis-Cummings Hamiltonian with disorder
has been the subject of a few recent papers \cite{Botzung2020,Du2021,Zhao2022JCP,Houdre1996,Gera2022} which give approximate ways of analyzing the problem.

Scholes \cite{Scholes2020} in an interesting paper studied excitation
energy transfer in a finite collection of molecules arranged in different
topologies, varying from linear to the star. A H$\ddot{u}$ckel type
Hamiltonian with nearest neighbor hopping was used as the model.  The effect of diagonal disorder
on the spectrum of the eigenvalues as well as the inverse participation
ratio was investigated. The diagonal disorder on the sites were taken
to be independent identically distributed Gaussian random variables and all the quantities
were numerically calculated. Among all topologies considered, the
star graph (star topology) was found to have the most stable eigenvalue spectrum.  Very interestingly, the Tavis-Cummings Hamiltonian has the
same topology as the star graph. 

In this paper we point out that it is possible to find {\em exact results} for
an arbitrary one electron Hamiltonian with diagonal disorder, provided
the disorder has a Cauchy distribution. The disorder numerically studied
by Scholes is one in which the energy of a given site $``i"$ is given
by $\epsilon_{a}+\xi_{i}$ with $\xi_i$ following the Gaussian distribution.
In comparison, the Cauchy or Lorentz distribution is given by $P(\xi_{i})=\frac{\gamma}{\pi(\gamma^{2}+\xi_{i}^{2})}$. 
 It has been realized long ago by
Lloyd \cite{Lloyd1969} that some exact results can be calculated
analytically for tight binding model Hamiltonians with site disorder
having a Cauchy distribution. %Lloyd proved this analytically using
%a recursive Green's function method. It is possible to do the same
%thing using the replica trick, but
 An elegant method to show this is the use
of Grassmann variables \cite{Haake2010,Zinn-Justin-PI}. We give the barest
minimum details of the method. 

 We consider a one electron Hamiltonian which may be written as 
$
\hat{H}=\sum_{i,j}^{N}H_{ij}|i\rangle\langle j|,\label{eq:Hamiltonian}
$
with diagonal disorder.  $N$ is the number of sites, each having an orbital $\left|i\right\rangle $.
$H_{ij}$ represents the hopping matrix element between the $i^{th}$
and $j^{th}$ sites. The site energy  $H_{ii}$ has disorder and is given by 
$
H_{ii}=\epsilon_{a}+\xi_{i},
$
where $\xi_{i}$ is the random component,  having the
Cauchy probability distribution $P(\xi_i)$.
 We shall use the symbol $\boldsymbol{H}$ to
denote the matrix with matrix elements $H_{ij}$.  We denote by $<..>_{\boldsymbol{\xi}}$
the average over all $\xi_{i}$s and clearly, the mean value $<\xi_{i}>_{\boldsymbol{\xi}}=0$.
 We are interested in calculating the Green's matrix
$
\boldsymbol{G}(\omega)=\left (\omega-\boldsymbol{H}\right )^{-1}$ and its average $<\boldsymbol{G}(\omega)>_{\boldsymbol{\xi}}$.
The matrix elements of $\boldsymbol{G}(\omega)$ are very useful for
the calculation of the one-particle properties of the system. For
example, defining the eigenstates $|m\rangle$ of the Hamiltonian %(\ref{eq:Hamiltonian})
by 
$
\hat{H}|m\rangle=\varepsilon_{m}|m\rangle
$
we can calculate $\rho_{i}(\omega)$ the one particle density of states
on the $i^{th}$ orbital $|i\rangle$. 
$
\rho_{i}(\omega)  =\sum_{m}\left|\left\langle i\left|m\right.\right\rangle \right|^{2}\delta(\omega-\varepsilon_{m})
  =-\frac{1}{\pi}Im\left[G_{ii}(\omega+i\eta)\right]$,
 with $\eta$ being positive and infinitesimal.   The quantity that is of greatest interest is the
average of the density of states on the $i^{th}$ site $\left\langle \rho_{i}(\omega)\right\rangle _{\boldsymbol{\xi}},$
which implies that we need to calculate $\left\langle G_{ii}(\omega+i\eta)\right\rangle _{\boldsymbol{\xi}}$.
%\section{Calculation of $\left\langle G_{ii}(\omega+i\eta)\right\rangle _{\boldsymbol{\xi}}$ }
%\label{sec:Calculation-of-G-ii}
%
%We now look into the problem of calculating the $\left\langle G_{ii}(\omega+i\eta)\right\rangle _{\boldsymbol{\xi}}.$
For this we follow  Zinn-Justin \cite{Zinn-Justin-PI}
and define a ``generating function", 
$ %\begin{equation}
\mathfrak{Z}(\boldsymbol{A},\boldsymbol{b},\overline{\boldsymbol{b}})=\int\left(\prod_{i=1}^{N}\frac{dz_{i}d\overline{z}_{i}}{2\pi i}\right)\exp\left(-\overline{\boldsymbol{z}}.\boldsymbol{A}.\boldsymbol{z}+\overline{\boldsymbol{b}}.\boldsymbol{z}+\overline{\boldsymbol{z}}.\boldsymbol{b}\right),\label{eq:generating-function}
$%\end{equation}
 where $\boldsymbol{z}=(z_{1,}z_{2},z_{3},...z_{N})^{T}$, $z_{i}$
being a complex variable and the $\boldsymbol{\overline{z}}=(\overline{z}_{1},\overline{z}_{2},...\overline{z}_{N}),$ where ``bar" is used to denote the complex conjugate. 
$\boldsymbol{b}=(b_{1,}b_{2},b_{3},..b_{N})^{T}$ and $\boldsymbol{\overline{b}}=(\overline{b}_{1},\overline{b}_{2},..\overline{b}_{N})$
with $b_{i}$ arbitrary complex variables. The superscript $T$ is
used to denote transpose of a matrix and $
\overline{\boldsymbol{z}}.\boldsymbol{A}.\boldsymbol{z}=\sum_{i,j=1}^{N}\bar{z}_{i}A_{ij}z_{j},
$
where $\boldsymbol{A}$ is the matrix having $A_{ij}$ as its matrix
elements. Zinn-Justin \cite{Zinn-Justin-PI} shows that 
$
\mathfrak{Z}(\boldsymbol{A},\boldsymbol{b},\overline{\boldsymbol{b}})=|\boldsymbol{A}|^{-1}\exp\left(\overline{\boldsymbol{b}}.\boldsymbol{A^{-1}}.\boldsymbol{b}\right),
$
and gives the conditions under which this is true. Here $|\boldsymbol{A}|$ stands for the determinant of the matrix $\boldsymbol{A}$. This can be rearranged to get 
$
\left[\frac{\partial^{2}}{\partial \overline{b}_{i}\partial b_{j}}\mathfrak{Z}(\boldsymbol{A},\boldsymbol{b},\overline{\boldsymbol{b}})\right]_{\boldsymbol{b}=\boldsymbol{0}}=|\boldsymbol{A}|^{-1}\left(\boldsymbol{A}^{-1}\right)_{ij},\label{eq:A-1ii}$
\begin{equation}
\left(\boldsymbol{A}^{-1}\right)_{ij}=\left[|\boldsymbol{A}|\frac{\partial^{2}}{\partial \overline{b}_{i}\partial b_{j}}\mathfrak{Z}(\boldsymbol{A},\boldsymbol{b},\overline{\boldsymbol{b}})\right]_{\boldsymbol{b}=\boldsymbol{0}}.\label{eq:A-invii}
\end{equation}
 For calculating $G_{ij}(\omega+i\eta)$ we put
$\boldsymbol{A}=$$-i(\omega+i\eta-\boldsymbol{H)}$ . The quantity
$\eta$  ensures that
the integral in the definition of $\mathfrak{Z}(\boldsymbol{A},\boldsymbol{b},\overline{\boldsymbol{b}})$ is convergent.
From Eq. (\ref{eq:A-invii}), we find  
\begin{eqnarray}
\left\langle \left[(\omega+i\eta-\boldsymbol{H)}^{-1}\right]_{ii}\right\rangle _{\boldsymbol{\xi}}= \left[\frac{\partial^{2}}{\partial \overline{b}_{i}\partial b_{j}}\left\langle \det |-i(\omega+i\eta-\boldsymbol{H)}|\mathfrak{\quad Z}({-i(\omega+i\eta-\boldsymbol{H)}},\boldsymbol{b},\overline{\boldsymbol{b}})\right\rangle _{\boldsymbol{\xi}}\right]_{\boldsymbol{b}=\boldsymbol{0}}.\label{eq:G_as_product}
\end{eqnarray}
In our case $\boldsymbol{H}$ has the random component $\boldsymbol{\xi}$
which needs to be averaged over $\boldsymbol{\xi}$ as indicated on the right hand side
of the above equation. This averaging seems impossible, because it
is a product of the determinant and the quantity $\mathfrak{\quad Z}({-i(\omega+i\eta-\boldsymbol{H)}},\boldsymbol{b},\overline{\boldsymbol{b}})$ both
of them dependent on $\boldsymbol{\xi}.$  Interestingly, it
is possible to write the determinant in a form such that the averaging
can be done easily using Grassmann variables \cite{Zinn-Justin-PI,Haake2010}.
We note that 
$
 |-i(\omega+i\eta-\boldsymbol{H)}|=\int d\overline{\boldsymbol{\rho}}d\boldsymbol{\rho}e^{i\overline{\boldsymbol{\rho}}(\omega+i\eta-\boldsymbol{H)\boldsymbol{\rho}}},\label{eq:Grassmann-for-determinant}
$
where $\boldsymbol{\rho}=\left(\rho_{1},\rho_{2},...\rho_{N}\right)^T$
and $\overline{\boldsymbol{\rho}}=\left(\overline{\rho}_{1},\overline{\rho}_{2},...\overline{\rho}_{N}\right)$
are both collections of $N$ Grassmann variables, $\bar{\rho_{i}}$
being the complex conjugate of $\rho_{i}.$ Hence, 
\begin{equation}
\left\langle 
\left[
(\omega+i\eta-\boldsymbol{H)}^{-1} \right]_{ii}
\right\rangle_{\boldsymbol{\xi}}
=\left[
\frac{\partial^{2}}{\partial b_{i}^{*}\partial b_{j}}
\int d\boldsymbol{z}d\overline{\boldsymbol{z}} 
\int d\overline{\boldsymbol{\rho}}d\boldsymbol{\rho}\left\langle 
e^{
i\overline{\boldsymbol{\rho}}(\omega+i\eta-\boldsymbol{H})\boldsymbol{\rho}+
i\overline{\boldsymbol{z}}
(\omega+i\eta -\boldsymbol{H})\boldsymbol{z}
+\overline{\boldsymbol{b}}.
\boldsymbol{z}+\overline{\boldsymbol{z}}.
\boldsymbol{b} 
}\right\rangle_{\boldsymbol{\xi}}
\right ]_{\boldsymbol{b}=\boldsymbol{0}}
.\label{eq:G-as-superintegral}
\end{equation}

The integral in the above is over ordinary variables as well as the
anti-commuting Grassmann variables and is referred to as a
superintegral and   the approach itself  as the supersymmetric
method. As the $\xi_{j}$s in Eq. (\ref{eq:G-as-superintegral})  appears in the exponential, one can
average each $\xi_{j}$ separately. The result of collecting together
just the terms that depend on $\xi_{j}$ and then averaging is demonstrated
below:
\begin{eqnarray}
\left\langle e^{-i\xi_{j}(\rho_{j}^{*}\rho_{j}+z_{j}^{*}z_{j})}\right\rangle _{\boldsymbol{\xi}} & = & \frac{\gamma}{\pi}\int_{-\infty}^{\infty}d\xi_{j}\frac{\exp\left\{ i\xi_{j}(\rho_{j}^{*}\rho_{j}+z_{j}^{*}z_{j})\right\} }{(\gamma^{2}+\xi_{j}^{2})}=
  e^{ -\gamma(\rho_{j}^{*}\rho_{j}+z_{j}^{*}z_{j})} .\label{eq:averaging-result}
\end{eqnarray}
Using Eq. (\ref{eq:averaging-result}) in Eq. (\ref{eq:G-as-superintegral})
we get

$ %\begin{equation}
\left\langle 
\left[
{(\omega+i\eta -\boldsymbol{H})}^{-1}
\right  ]_{ii}
\right\rangle_{\boldsymbol{\xi}}
=\left[
\frac{\partial^{2}}{\partial b_{i}^{*}\partial b_{j}}
\int d\boldsymbol{z}d\overline{\boldsymbol{z}}
\int d\overline{\boldsymbol{\rho}}d\boldsymbol{\rho}
e^{i\overline{\boldsymbol{\rho}}
(\omega-\boldsymbol{H}_{0}+i\gamma \boldsymbol{I})\boldsymbol{\rho}
+i\overline{\boldsymbol{z}}
(\omega-\boldsymbol{H}_{0}+i\gamma \boldsymbol{I})\boldsymbol{z}+\overline{\boldsymbol{b}}.\boldsymbol{z}+\overline{\boldsymbol{z}}.\boldsymbol{b}}\right]_{\boldsymbol{b}=\boldsymbol{0}},\label{eq:Averaged-Gii}\\
$ %\end{equation}
which simplifies the calculation enormously.
It shows that for the purpose of the calculation of the averaged quantity,
the random Hamiltonian can be replaced with a complex one in which
the random term $\xi_{i}$ is simply replaced by the non-random  imaginary quantity $-i\gamma$.
Thus one has the complex, non-Hermitian Hamiltonian 
$ %\begin{equation}
\hat{\cal H} =\sum_{i}^{N}(\epsilon_{a}-i\gamma)|i\rangle\langle i|+\sum_{i\neq j}^{N}\left\{ H_{ij}|i\rangle\langle i|+H_{ji}|i\rangle\langle i|\right\} ,\label{eq:complex-hamiltonian}
$ %\end{equation}
for which  we need to calculate the Green's matrix.
  From this the averaged density of states on the $i^{th}$ site
may be easily calculated as 
$
\langle\rho_{ii}(\omega)\rangle_{\boldsymbol{\xi}}=-\frac{1}{\pi}Im\langle i|\left(\omega-\hat{\cal H} \right)^{-1}|i\rangle.
$
We now proceed to illustrate the utility of the approach for two Hamiltonians  with diagonal disorder.  They are the Tavis-Cummings Hamiltonian for molecules confined to a microcavity and the H$\ddot{u}$ckel type Hamiltonian for describing exciton transfer.

%\section{Solution of the disordered Tavis-Cummings model }

\label{sec:Exact-solution-Tavis-Cummings-1}A problem of great current
interest is the different excited states of a large number of molecules
put inside a cavity of size such that a molecular excitation is nearly
resonant with a standing mode of the cavity. This is modelled by the
Tavis-Cummings Hamiltonian for which exact solution is easy to find.
However, there are only approximate analyses for the case where the
molecular excitation energies are  randomly distributed around a mean. Recently,
we have obtained an approximate analytic solution for the case the excitation
energies have a Gaussian distribution \cite{Gera2022}. The solution is
expected to asymptotically approach the exact solution as the number
of molecules $N\rightarrow\infty.$ Using the method outlined above, we can find the analytic solution
any value of $N$, small or large. In our model,  the excitation energy $\epsilon_{i}$ of the $i^{th}$
molecule is taken as 
$
\epsilon_{i}=\epsilon_{a}+\xi_{i},\label{Eq.epsilon1-1}
$
where $\xi_{i}$s are identically distributed random variables having
the Cauchy probability distribution.

With such a disorder, the Hamiltonian may be written as 
$ %\begin{equation}
\hat{H}=\epsilon_{c}|c\rangle\langle c|+\sum_{i}^{N}(\epsilon_{a}+\xi_{i})|i\rangle\langle i|+\sum_{i}^{N}(V_{i}|c\rangle\langle i|+V_{i}^{*}|i\rangle\langle c|).
$ %\end{equation}
$|c\rangle$ denotes the state where the cavity mode is excited to
its first excited level. It has an energy of $\epsilon_{c}$. $|i\rangle$
denotes the state in which the $i^{th}$ molecule is excited. This problem has $N+1$ states, one from the cavity and one each from the $N$ molecules.  The
coupling constant of the cavity mode to the excited state of the $i^{th}$ molecule  is
denoted by $V_{i}$  and is given by 
$
V_{i}=-\boldsymbol{E}(z_i).\boldsymbol{{\mu}}_i=-\mu_{eg}\sqrt{\frac{\epsilon_{c}}{2\epsilon_{o}{\cal V}}}=V,\label{cc}
$
where we have assumed that all the molecules are oriented in the direction of the electric field $\boldsymbol{E}(z_i)$ of the cavity and that the positional variation  of the electric field may be neglected.  ${\cal V}$ is the volume of the cavity and $\boldsymbol{{\mu}}_i$ is transition dipole moment of the molecular excitation, having  the magnitude  $\mu_{eg}$.  The permittivity of matter and that of free space are denoted as  $\epsilon_c$ and $\epsilon_{o}$ respectively.   In the following,  letters like $G,\boldsymbol{G}, \hat{H},...$ are quantities that depend on the disorder $\boldsymbol{\xi}$. Symbols like $\cal{G},\boldsymbol{{\cal G}},\boldsymbol{{\cal H}}, .....$ are quantities that have been averaged over $\boldsymbol{\xi}$. 

 To calculate 
$
\left\langle G_{ij}(\omega)\right\rangle _{\boldsymbol{\xi}}=\langle i|\left[\omega-\hat{\cal H} \right]^{-1}|j\rangle,
$
we need to use the complex Hamiltonian with all the random terms replaced with $-i\gamma$.  This gives the new Hamiltonian $
\hat{\cal H} =\epsilon_{c}|c\rangle\langle c|+\sum_{i}^{N}(\epsilon_{a}-i\gamma)|i\rangle\langle i|+\sum_{i}^{N}(V_{i}|c\rangle\langle i|+V_{i}|i\rangle\langle c|).
$
We now proceed to determine the matrix elements of $\hat{\cal G} (\omega)=(\omega-\hat{\cal H} )^{-1}$
following \cite{Gera2022} and find 
$
{\cal G}_{cc} (\omega)=\left[\omega-\epsilon_{c}-{\Sigma}(\omega)\right]^{-1},\label{Eq.Gcc1-1}
$
with the self energy  
$
\Sigma(\omega)=\sum_{i=1}^N \frac{V_i^2}{\omega-\epsilon_{a}+i\gamma} =\frac{NV^2}{\omega-\epsilon_{a}+i\gamma}=\frac{{\cal N}\tilde{V}^2}{\omega-\epsilon_{a}+i\gamma}.
$ 
 We have defined $\tilde{V}=\sqrt{\cal{V}}V$ and the number density of molecules within the cavity ${\cal N}=N/\cal{V}$. See the Supporting Information for more details. Then
$
{\cal G}_{cc} (\omega)  =  \left[\omega-\epsilon_{c}-\frac{{\cal N}\tilde{V}^{2}}{\omega-\epsilon_{a}+i\gamma}\right]^{-1}.
$
 The poles of ${\cal G}_{cc} (\omega)$ are at $
\epsilon_{\pm}=\frac{\left(\epsilon_{a}+\epsilon_{c}-i\gamma\right)}{2}\pm\sqrt{{\cal N}\tilde{V}^{2}+\left(\frac{\epsilon_{c}-\epsilon_{a}-i\gamma}{2}\right)^{2}}.
$
The real part of $\epsilon_\pm$ determine the energies of the two polaritonic states and the imaginary part its lifetime.
% $G_{cc}(\omega)$ can be written as: 
%\begin{eqnarray}
%G_{cc}(\omega) & = & \frac{\epsilon_{a}-i\gamma-\omega}{{\cal N}\tilde{V}^{2}-(\omega-\epsilon_{c})(\omega-\epsilon_{a}+i\gamma)}.
%\end{eqnarray}
%\[
%\rho_{T}(\omega)=\frac{1}{\pi}\frac{NV^{2}\gamma}{\gamma^{2}(\epsilon_{c}-\omega)^{2}+(NV^{2}+(\epsilon_{a}-\omega)(\omega-\epsilon_{c}))^{2}}-\frac{1}{\pi}Im\{\frac{1}{V^{2}}\Sigma(\omega^{+})-\frac{\partial\Sigma(\omega^{+})}{\partial\omega}G_{cc}(\omega^{+})\}
%\]

%\subsection{Width of the polaritonic peaks}

The widths of the peaks are determined by the imaginary part of $\epsilon_\pm$.  In the limit of large $\sqrt{\cal{N}}|\tilde{V}|$ it is found to be $-i\gamma/2$. This means that the two polaritons have the same width, which is equal to half the width of the distribution of the molecular states.  We note that the widths of the polaritonic peaks are very sensitive to the nature of the distribution of molecular energy levels, i.e.,
the probability distribution $P(\xi_{i})$. In our previous work \cite{Gera2022} we
had analyzed the case of Gaussian distribution and found that the
width decreases as the value of $\cal N$ becomes large. In the case of Cauchy distribution, 
 the width remains constant provided $ \sqrt{\cal{N}}|\tilde{V}|\gg \gamma$. On the other hand, if $P(\xi_{i})$
is taken as a uniform distribution in the range $(-w,w),$ then it
is found that the polarionic peaks are Dirac delta functions.
%\subsection{The Absorption spectrum}
The absorption cross-section
is given by \cite{Gera2022}: 
$
\alpha(\omega)  =-\frac{\omega}{\epsilon_{o}c\hbar}N\left|\boldsymbol{{\mu}}\right|^{2}\;Im\left[{\cal G}_{mol,mol}(\omega)\right]\label{eq:Eq.CS-1}
$
where, 
$
{\cal G}_{mol,mol}(\omega)  =  (\omega-\epsilon_{c}){\cal {G}}_{cc}(\omega)\frac{1}{\mathscr{N}\tilde{V}^{2}}\Sigma(\omega).
$
Derivations for this as well as other quantities may be found in the Supporting Information.

\begin{figure}[h]
\includegraphics[width=0.8\linewidth]{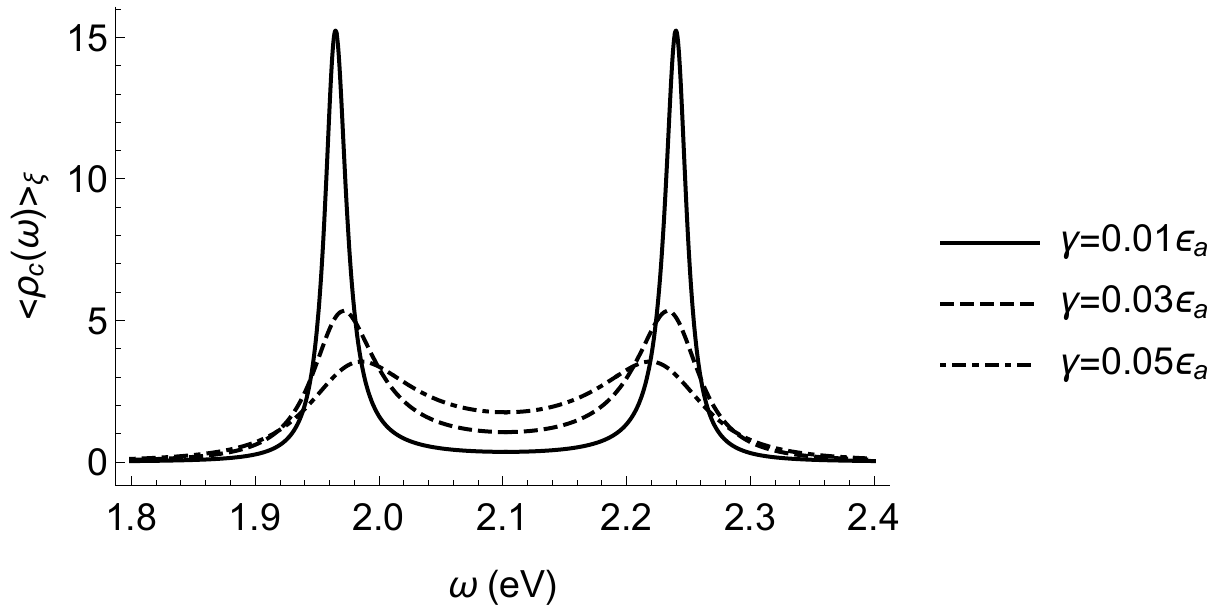} \caption{Plot for $<\rho_{c}(\omega)>_{\boldsymbol{\xi}}$ against $\omega$ for $\mathscr{N}=1.16\times10^{25}m^{-3}$
and different values of $\gamma$. Following values were used $\epsilon_{c}=2.1\,eV$,
$\epsilon_{a}=\epsilon_{c}$, $\tilde{V}=4.06\times10^{-14}\, eVm^{3/2}$.  }
\label{rhocc-1} 
\end{figure}

In Fig. {\ref{rhocc-1}} the cavity density of states is plotted
against $\omega$ with varying values of $\gamma$. The cavity state
is divided amongst the two polaritonic states with total area under
the two peaks equals to unity. As the value of $\gamma$ increases the
peak width increases as we would expect. 

%\subsection{Change in density of states\label{subsec:Change-density-of-1}}

\begin{figure}[h]
\includegraphics[width=0.8\linewidth]{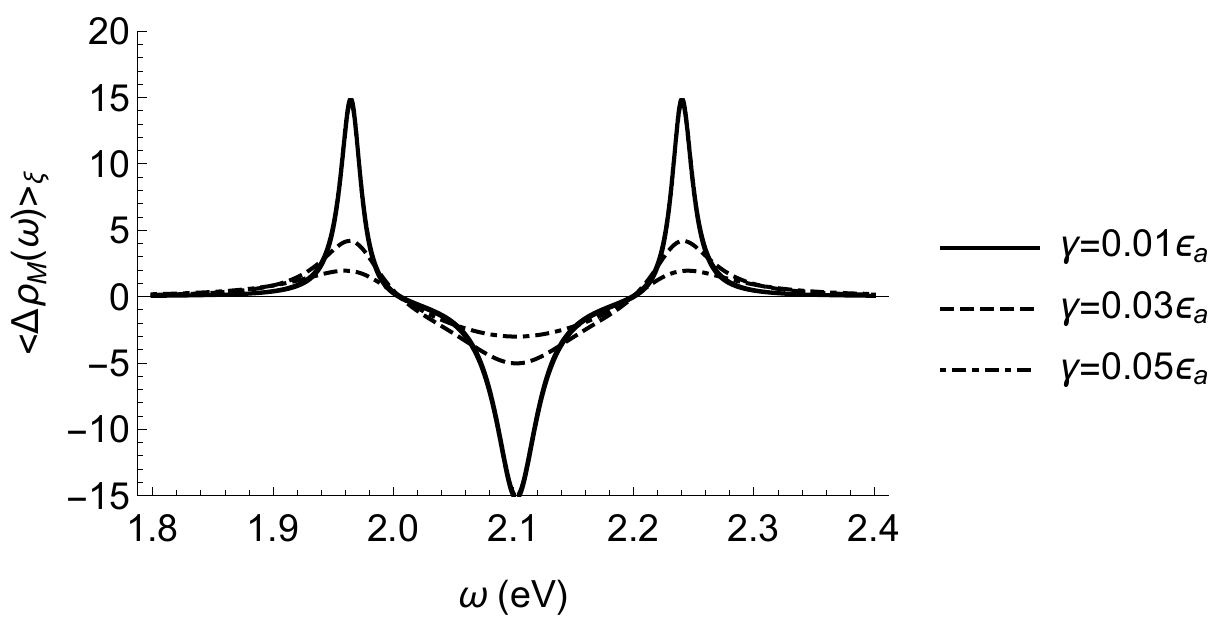} \caption{Plot for $<\Delta \rho_{M}(\omega)>_{\boldsymbol{\xi}}$ against $\omega$ for $\mathscr{N}=1.16\times10^{25}m^{-3}$
and different values of $\gamma$. Following values were used $\epsilon_{c}=2.1\,eV$,
$\epsilon_{a}=\epsilon_{c}$, $\tilde{V}=4.06\times10^{-14}\,eVm^{3/2}$. }
\label{drhoc} 
\end{figure}

In Fig. {\ref{drhoc}} the change in molecular density of states $\Delta\rho_{M}(\omega)$ (expression given in Supporting Information),
 $\omega$ with varying values of $\gamma$.  
Integrating the area under the curve from $\omega=2.0\;eV$ to $\omega=2.2\;eV$
results in the value of $-1$. This means that out of the $N$ states, effectively one
state contributes to the polariton formation.  Further, the dip in the middle means that  the states which
are in close resonance to the cavity state contributes the largest to the polaritons.   

%\subsection{Absorption Spectrum}

%\label{subsec:Absorption-Spectrum-1}

\begin{figure}[h]
\includegraphics[width=0.8\linewidth]{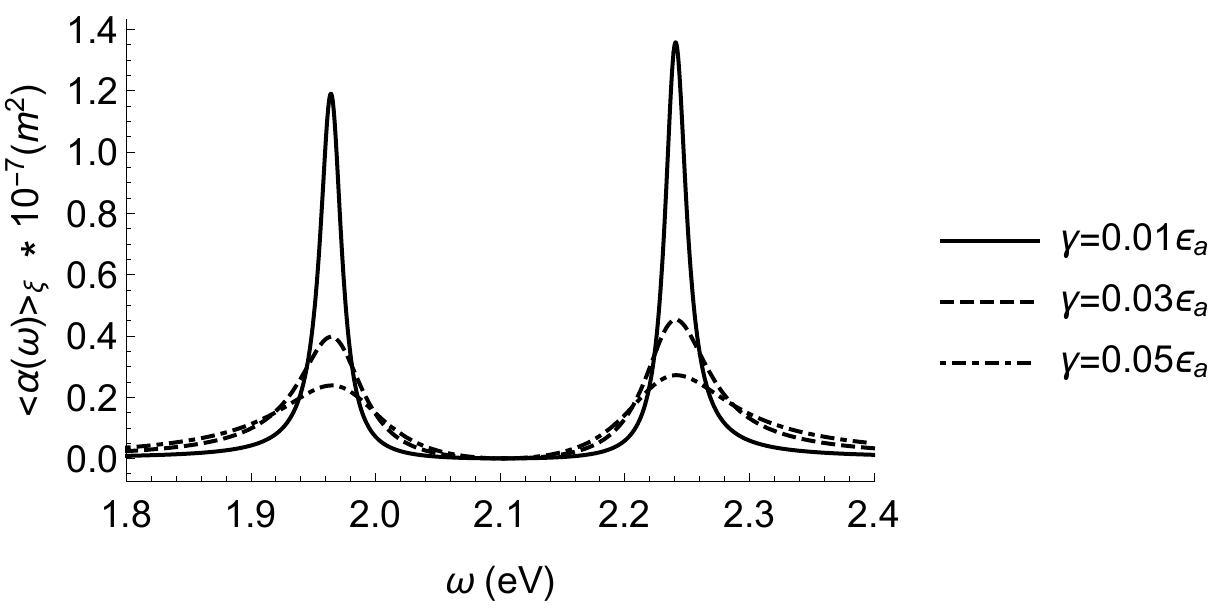} \caption{Plot for absorption cross section $<\alpha (\omega)>_{\boldsymbol{\xi}}$ against $\omega$ for $\mathscr{N}=1.16\times10^{25}m^{-3}$
and varied values of $\gamma$. Following values were used $\epsilon_{c}=2.1\,eV$,
$\epsilon_{a}=\epsilon_{c}$, $\tilde{V}=4.06\times10^{-14}\,eVm^{3/2}$, $|\boldsymbol{\mu}|=10D$.  }
\label{FigCS-1} 
\end{figure}

It is possible to calculate  absorption spectrum of the system using expressions given in Supporting Information, for varied values of $\gamma$ and the results 
are shown in Fig. \ref{FigCS-1}. 

%\newpage
%\newpage
%\section{H$\ddot{u}$ckel type hamiltonians with diagonal disorder - exact
%results}

\label{sec:Huckel-type-hamiltonians} We now consider the second Hamiltonian with disorder.  In connection with the cavity problem, Scholes \cite{Scholes2020} posed the question: how to  maximize delocalization of excitons among many sites in presence of disorder?  To answer this question he did numerical investigation of different topologies of arranging the sites.  The investigation involved generating many copies of the same system with randomly distributed site energies obeying a Gaussian distribution  and studying the  properties of eigenstates and eigenvalues averaged over the different realizations.  Inspired by this,  we present exact analytical results for the same kind of problem, but with Cauchy disorder. 
 
   There are $N$ sites in the model which are  
 molecules, each of which has
one electronic excitation, that can hop from one site to the next. The
on-site excitation energy has the form $\alpha+\xi_{i}$ for the $i^{th}$
site. Here $\xi_{i}$ is the random component, obeying Cauchy distribution.
The hopping happens only if the two molecules are connected (neighbors)
and the matrix element for this is denoted as $\beta$. Without loss of generality, we take $\alpha=0$ and $\beta=1$.  For such a problem with $N$ sites arranged linearly, the averaged Green's function matrix elements $\left\langle \boldsymbol{G}(\omega)\right\rangle _{\boldsymbol{\xi}}$  may be calculated as $(\omega -{\boldsymbol{ \cal H}})^{-1}$ using the complex Hamiltonian   
\begin{equation}
\boldsymbol{\cal H} =\left[\begin{array}{ccccc}
-i\gamma & 1 & ... & 0 & 0\\
1 & -i\gamma & ... & .. & 0\\
.. & .. & .. & .. & ..\\
0 & .. & .. & -i\gamma & 1\\
0 & 0 & .. & 1 & -i\gamma
\end{array}\right].
\end{equation}
 We note that 
$
\boldsymbol{\cal H} =\boldsymbol{H}_{0}-i\gamma\boldsymbol{I},
$
where $\boldsymbol{I}$ is the $N\times N$ identity matrix. Here $\boldsymbol{H}_{0}=\boldsymbol{\cal{H}}(\gamma=0)$. Let us denote
by $\boldsymbol{U}$ the unitary matrix that diagonalises $\boldsymbol{H}_{0}$.
i.e., 
$
\boldsymbol{U}^{\dagger}{\boldsymbol{H}_{0}}\boldsymbol{U}=\boldsymbol{\varepsilon},
$
where $\boldsymbol{\varepsilon}$ is the diagonal matrix having the eigenvalues
$\varepsilon_{m}$, $m=1,2,3,....N$ of $\boldsymbol{H}_{0}$ as its
diagonal elements. It is clear that the same $\boldsymbol{U}$ diagonalizes
$\boldsymbol{\cal H} $. The eigenvalues of $\boldsymbol{\cal H} $
are given by the diagonal matrix $\boldsymbol{\varepsilon}-i\gamma\boldsymbol{I}$.
Also, 
$%\begin{align}
\left\langle \boldsymbol{G}(\omega)\right\rangle _{\boldsymbol{\xi}}  =\boldsymbol{\cal G}(\omega)=\left[\omega I-\boldsymbol{\cal H} \right]^{-1} 
  =\boldsymbol{U}^{\dagger}\left[\omega I-\boldsymbol{U}^{\dagger}\boldsymbol{\cal H} \boldsymbol{U}\right]^{-1}\boldsymbol{U}
  =\boldsymbol{U}^{\dagger}\left[\left(\omega+i\gamma\right)I-\boldsymbol{\boldsymbol{\varepsilon}}\right]^{-1}\boldsymbol{U}.\label{eq:<G>}
$ 
 This result shows that one needs only to calculate the eigenvectors
($\boldsymbol{U}$) and eigenvalues $\boldsymbol{\boldsymbol{\varepsilon}}$
of the Hamiltonian without disorder. Then the average value $\left\langle \boldsymbol{G}(\omega)\right\rangle_{\boldsymbol{\xi}} $
may be easily calculated using the above expression.

\begin{figure}[h]
	\includegraphics[width=0.5\linewidth]{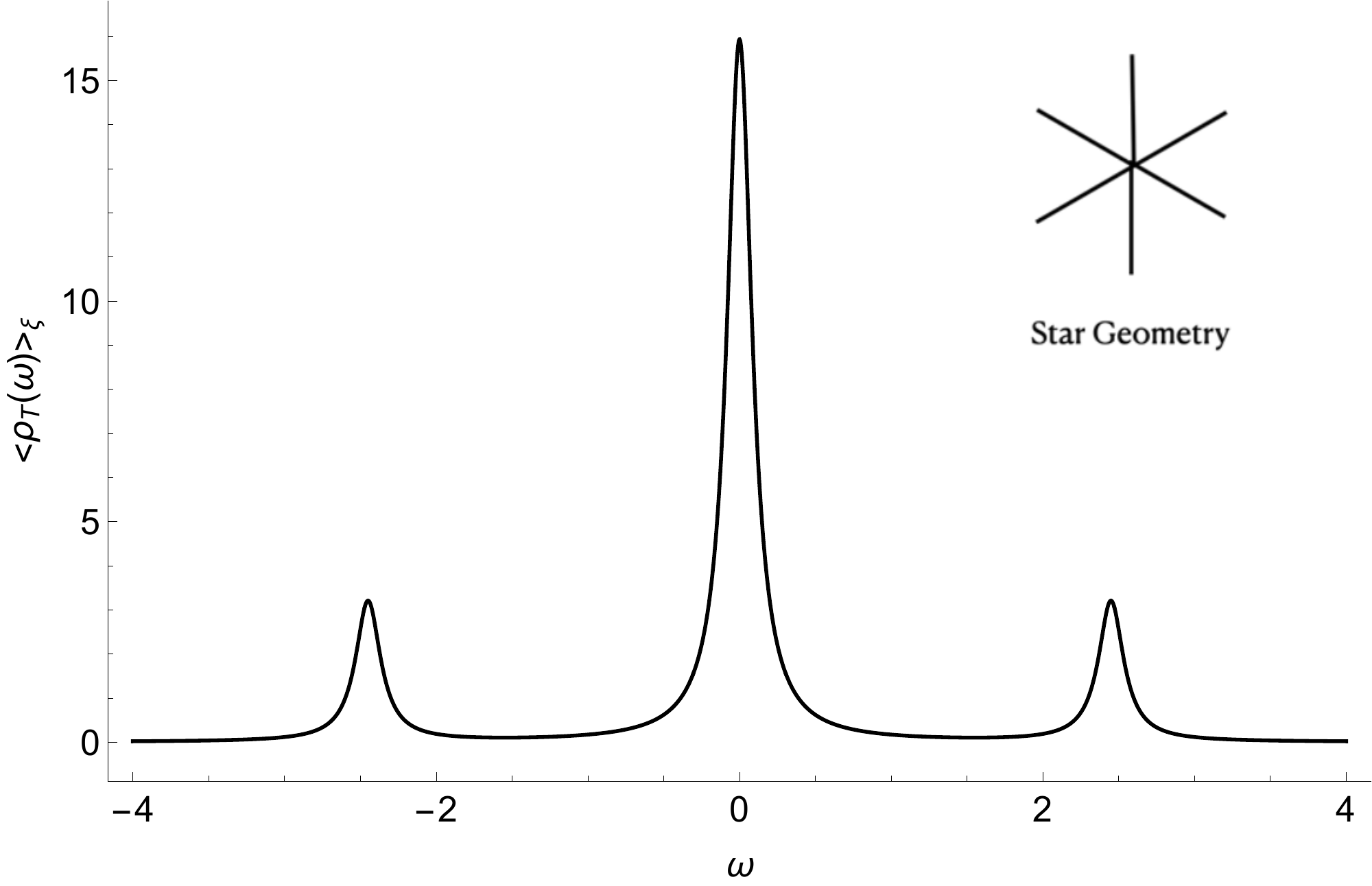} \caption{Plot of $\rho_{T}(\omega)$ against $\omega$ for star topology with  7 sites and $\gamma=0.1$. The central peak is 5 times more intense than each of the side peaks.  \label{FigrhoS7} }
\end{figure}

 To illustrate the method, we report the averaged total density of states for an arrangement of seven sites with star topology, in  Fig. \ref{FigrhoS7}. The topology is shown as an inset in the same figure.  The topology of this is the same as the one in the cavity problem, with the star having $N$ (=total number of molecules) arms. The density of states for the star topology with seven sites is shown in Fig. \ref{FigrhoS7}. Just as in the case of cavity problem, the resultant density of states has a central peak similar to dark states and two polariton like peaks on the two sides of the central peak.
In the Supporting Information, we give results for a hexagonal unit of six sites. 

%\section{Conclusions}
In summary, we have shown that it is possible to find exact results for arbitrary one electron Hamiltonians with diagonal disorder, if the disorder has a Cauchy distribution.  Using supersymmetric techniques, we showed that the the random terms in the Hamiltonian may be replaced by the non-random term $-i\gamma$ and the resultant complex Hamiltonian may be used to calculate averaged one-electron properties.  Also, we showed that it is enough to diagonalize and find the eigenvalues of the Hamiltonian without any disorder, and then the results for all one particle properties for the disordered system may  be obtained, avoiding the usual method of generating multiple realizations of the system, calculating the properties for each and then averaging them to get the averaged properties.   Using the method,  we  report exact results for the disordered Tavis-Cummings Hamiltonian with arbitrary number of molecules.  The solution shows the existence of two polaritonic states having a width which is half that of the distribution of the site energy on a given molecule.  We pointed out that the width of the polaritonic peaks sensitively depend on the way the site energy is distributed.  For a Gaussian distribution, the peaks narrow as the splitting increases, while for Cauchy is remains constant.  As typical examples of other problems that could be solved exactly, we have also presented exact solutions for H$\ddot{u}$ckel type Hamiltonian, for an arrangement of sites having star topology.   It is possible to use the method of this paper for any collection of sites with arbitrary topology. An interesting problem that remains is the analytical calculation of the inverse participation ratio (IPR).  This requires calculation of the average of a product of two Green's functions, and though the averaging can be done using the supersymmetric method, further progress seems difficult.  
\bibliography{ms}
%\bibliographystyle{unsrt}
%\newpage

\end{document}

\begin{center}
{\Large \bf Supplementary Information} \end{center}

\subsection{Derivation of the results for the cavity problem}
We remind the reader that $ NV^2 ={\cal{N}}\tilde{V}^2$, where $\cal{N}$ is the number density of the molecules and $\tilde{V}=-\mu_{eg}\sqrt{\frac{\epsilon_c}{2\epsilon_o }}$.
As discussed in the paper, the average of the Green's function $\left\langle \boldsymbol{G}\right\rangle _{\boldsymbol{\xi}}$
can be easily calculated using the result $\left\langle \boldsymbol{
{ G}}\right\rangle _{\boldsymbol{\xi}}=\mathbf{{ \boldsymbol{\cal G}}}=\left(\omega-\boldsymbol{{\cal H}}\right)^{-1}$
where $\boldsymbol{{\cal H}}$ is the complex Hamiltonian. Explicitly,
\begin{equation}
\omega\boldsymbol{{\cal -H}}=\left[\begin{array}{c|ccccccc}
\omega-\epsilon_{c}\; & \;-V\; & \;-V\; & \;.\; & \;.\; & \;.\; & \;.\; & \;-V\;\\
\hline -V & \omega-\epsilon_{a}+i\gamma & 0 & . & . & . & . & 0\\
-V & 0 & \omega-\epsilon_{a}+i\gamma & . & . & . & . & 0\\
. & . & . & . & . & . & . & .\\
%.
-V & 0 & 0 & . & . & . & . & \omega-\epsilon_{a}+i\gamma
\end{array}\right]. \label{SIeq.1}
\end{equation}
As indicated in Eq.(\ref{SIeq.1}) this has the partitioned form 
\begin{equation}
\omega\boldsymbol{{\cal -H}}=\left[\begin{array}{c|c}
\omega-\epsilon_{c} & -\boldsymbol{V}\\
\hline -\boldsymbol{V}^{T} & \omega+i\gamma-\boldsymbol{\epsilon}_{M}
\end{array}\right].\label{Eq.H1}
\end{equation}
$M$ stands for molecules and $\boldsymbol{V}=[V,V,...,V]$ is a row
matrix having dimensions $1\times N$. $\boldsymbol{\epsilon}_M$ is an $N\times N$ diagonal matrix
of the excitation energies with randomness put equal to zero: 
\begin{equation}
\left(\boldsymbol{\epsilon}_{M}\right)_{ij}=\epsilon_{a}\delta_{ij}.
\end{equation}
$\boldsymbol{\mathbf{{\cal G}}}$ too can be written in the same partitioned
form as in Eq. (\ref{Eq.H1}): 
\begin{equation}
 \boldsymbol{\cal G}=\left[\begin{array}{c|c}
{\cal G}_{cc} & {\cal \boldsymbol{{\cal G}}}_{cM}\\
\hline \boldsymbol{{\cal G}}_{Mc} & {\cal \boldsymbol{{\cal G}}}_{MM}
\end{array}\right].\label{Eq.G2}
\end{equation}
Our analysis follows that of Anderson
\cite{Anderson1961} and Newns \cite{Newns1968}. $\boldsymbol{\cal G}$ obeys the equation 
\begin{equation}
(\omega\boldsymbol{I}-{\cal \boldsymbol{H}})\boldsymbol{\mathbf{{\cal G}}}=\boldsymbol{I}. \label{Eq.G1}
\end{equation}
Using Eq. (\ref{Eq.H1}),  Eq. (\ref{Eq.G2}) and Eq. (\ref{Eq.G1})
we get 
\begin{equation}
\left[\begin{array}{c|c}
\omega-\epsilon_{c} & -\boldsymbol{V}\\
\hline -\boldsymbol{V}^{T} & \omega+i\gamma-\boldsymbol{\epsilon}_{M}
\end{array}\right]\left[\begin{array}{c|c}
{\cal G}_{cc} & {\cal \boldsymbol{{\cal G}}}_{cM}\\
\hline {\cal \boldsymbol{{\cal G}}}_{Mc} & {\cal {\cal \boldsymbol{{\cal G}}}}_{MM}
\end{array}\right]=\boldsymbol{I}.\label{Eq.HG-1}
\end{equation}
Eq. (\ref{Eq.HG-1}) is equivalent to the following four equations:
\begin{eqnarray}
(\omega-\epsilon_{c}){\cal G}_{cc}-\boldsymbol{V}{\cal \boldsymbol{{\cal G}}}_{Mc} & = & 1\label{Eqset.1}\\
(\omega-\epsilon_{c})\boldsymbol{{\cal G}}_{cM}-\boldsymbol{V}\,\boldsymbol{{\cal G}}_{MM} & = & 0\\
-\boldsymbol{V}^{T}{\cal G}_{cc}+\left\{ \left(\omega+i\gamma\right)\boldsymbol{I}-\boldsymbol{\epsilon}_{M}\right\} \boldsymbol{{\cal G}}_{Mc} & = & 0\label{Eqset.2}\\
-\boldsymbol{V}^{T}\boldsymbol{{\cal G}}_{cM}+\left\{ \left(\omega+i\gamma\right)\boldsymbol{I}-\boldsymbol{\epsilon}_{M}\right\} {\cal \boldsymbol{{\cal G}}}_{MM} & = & \boldsymbol{I}.\label{Eqset.4}
\end{eqnarray}
Solving Eq. (\ref{Eqset.2}) for ${\cal \boldsymbol{\boldsymbol{{\cal G}}}}_{Mc}$
gives 
\begin{equation}
\boldsymbol{{\cal G}}_{Mc}=\left\{ \left(\omega+i\gamma\right)\boldsymbol{I}-\boldsymbol{\epsilon}_{M}\right\} ^{-1}\boldsymbol{V}^{T}\boldsymbol{{\cal G}}_{cc}.\label{Eq.Gcmdaggger}
\end{equation}
Using Eq. (\ref{Eq.Gcmdaggger}) in Eq. (\ref{Eqset.1}) and solving
for ${\cal G}_{cc}$ leads to 
\begin{equation}
{\cal G}_{cc}=\left[\omega-\epsilon_{c}-\Sigma(\omega)\right]^{-1},\label{Eq.Gcc1}
\end{equation}
with the self energy $\Sigma(\omega)$ defined by 
\begin{equation}
\Sigma(\omega)=\frac{\boldsymbol{VV}^{T}}{\omega+i\gamma-\epsilon_{a}}=\frac{NV^{2}}{\omega+i\gamma-\epsilon_{a}}=\frac{{\cal N}\tilde{V}^{2}}{\omega+i\gamma-\epsilon_{a}}.\label{Eq.SE-1}
\end{equation}

\subsection{The Density of States}

We define the averaged density of states for the cavity state and
calculate it from ${\cal G}_{cc}(\omega^{+})$ using 
\begin{eqnarray}
\left\langle \rho_{c}(\omega)\right\rangle _{\boldsymbol{\xi}} & = & \sum_{m=1}^{N+1}\left\langle \left|\langle c|m\rangle\right|^{2}\delta(\omega-\varepsilon_{m})\right\rangle _{\boldsymbol{\xi}}\nonumber \\
 & = & -\frac{1}{\pi}Im\{{\cal G}_{cc}(\omega)\}.
\end{eqnarray}
$\left\langle \rho_{c}(\omega)\right\rangle _{\boldsymbol{\xi}}$
gives us an idea of the amount of participation of the cavity state
in the $m^{th}$ eigenstate of the system. 

To calculate other quantities, we determine the matrix elements of
Green's operator for the molecular states. Eq. (\ref{Eqset.2}) can
be solved to get  
\begin{equation}
{\cal \boldsymbol{{\cal G}}}_{cM}=(\omega^+-\epsilon_{c})^{-1}\boldsymbol{V}{\cal \boldsymbol{{\cal G}}}_{MM}.
\end{equation}
Using this result in Eq. (\ref{Eqset.4}) gives 
\begin{equation}
{\cal \boldsymbol{{\cal G}}}_{MM}=\left[\left(\omega+i\gamma\right)\boldsymbol{I}-\boldsymbol{\epsilon}_{MM}-\frac{\boldsymbol{V}^{T}\boldsymbol{V}}{\omega^{+}-\epsilon_{c}}\right]^{-1}.\label{Eq.GM}
\end{equation}
The operator corresponding to the matrix $\boldsymbol{V}^{T}\boldsymbol{V}$
can be written in the form $\mathscr{N}\tilde{V}^{2}\left|mol\right\rangle \left\langle mol\right|$
with  
\begin{equation}
\left|mol\right\rangle =\frac{1}{\sqrt{N}}\sum_{i=1}^{N}\left|i\right\rangle .\label{eq:|mol>-definition}
\end{equation}
Hence we can write the operator corresponding to the matrix ${\cal \boldsymbol{{\cal G}}}_{MM}$
as 
\begin{equation}
\hat{{\cal G}}_{MM}(\omega)=\left[\left(\omega+i\gamma-\epsilon_{a}\right)\hat{I}-\frac{\mathscr{N}\tilde{V}^{2}\left|mol\right\rangle \left\langle mol\right|}{\omega^{+}-\epsilon_{c}}\right]^{-1}.
\end{equation}
Using the operator identity $(\hat{A}-\hat{B})^{-1}=\hat{A}^{-1}+\hat{A}^{-1}\hat{B}(\hat{A}-\hat{B})^{-1}$
valid for any two operators $\hat{A}$ and $\hat{B}$, we can write
\begin{align}
\hat{{\cal G}}_{MM}(\omega^{+}) & =\left(\omega+i\gamma-\epsilon_{a}\right)^{-1}\hat{I}+\left(\omega+i\gamma-\epsilon_{a}\right)^{-1}\frac{\mathscr{N}\tilde{V}^{2}\left|mol\right\rangle \left\langle mol\right|}{\omega^{+}-\epsilon_{c}}\hat{{\cal G}}_{MM}(\omega^{+}).\label{eq:G_M-1}
\end{align}
From this we find $\langle mol|\hat{{\cal G}}_{MM}(\omega^{+})$ to
obey the equation 
\begin{align}
\langle mol|\hat{{\cal G}}_{MM}(\omega^{+}) & =\left(\omega+i\gamma-\epsilon_{a}\right)^{-1}\langle mol|+\frac{\Sigma(\omega)}{(\omega^{+}-\epsilon_{c})}\langle mol|\hat{{\cal G}}_{MM}(\omega^{+}).
\end{align}
Solving for $\langle mol|\hat{{\cal G}}_{MM}(\omega^{+})$ gives 
\begin{align}
\langle mol|\hat{{\cal G}}_{MM}(\omega^{+})= & (\omega^{+}-\epsilon_{c}){\cal G}_{cc}(\omega^{+})\langle mol|\left(\omega+i\gamma-\epsilon_{a}\right)^{-1}.\label{eq:<mol|G_MM}
\end{align}
Using this back in Eq. (\ref{eq:G_M-1}) we get 
\begin{equation}
\hat{{\cal G}}_{MM}(\omega^{+})=\left(\omega+i\gamma-\epsilon_{a}\right)^{-1}\hat{I}+\left(\omega+i\gamma-\epsilon_{a}\right)^{-2}\langle mol|\mathscr{N}\tilde{V}^{2}\left|mol\right\rangle {\cal G}_{cc}(\omega^{+}).\label{eq:GM-2}
\end{equation}
Using Eq. (\ref{eq:<mol|G_MM}) the matrix element ${\cal G}_{mol,mol}(\omega^{+})=\left\langle mol\right|\hat{{\cal G}}_{MM}(\omega^{+})\left|mol\right\rangle $
can be calculated to be: 
\begin{eqnarray}
{\cal G}_{mol,mol}(\omega^{+}) & = & \frac{(\omega^{+}-\epsilon_{c}){\cal G}_{cc}(\omega)}{(\omega+i\gamma-\epsilon_{a})}.
\end{eqnarray}
From ${\cal G}_{mol,mol}(\omega^{+})$ we can calculate the density
of states of $|mol\rangle:$ 
\begin{equation}
<\rho_{mol}(\omega)>_{\boldsymbol{\xi}}=-\frac{1}{\pi}Im\{{\cal G}_{mol,mol}(\omega^{+})\}.\label{eq:rho_mol}
\end{equation}

The total density of states may be defined by $\rho_{T}(\omega)=\sum_{m=1}^{N+1}\delta(\omega-\varepsilon_{m})$
from which we get $<\rho_{T}(\omega)>_{\boldsymbol{\xi}}=-\frac{1}{\pi}Im\left(Tr{\hat{{\cal G}}(\omega^{+})}\right)$.
Using the complete set $|c\rangle,|1\rangle,|2\rangle,....|N\rangle$
to calculate the trace leads to 
\begin{equation}
<\rho_{T}(\omega)>_{\boldsymbol{\xi}}  =  <\rho_{c}(\omega)>_{\boldsymbol{\xi}} -\sum_{i=1}^{N}\frac{1}{\pi}Im\{{\cal G}_{ii}(\omega^{+})\},\label{Eq.rhot}
\end{equation}
with ${\cal G}_{ii}(\omega^{+})=\langle i|\hat{{\cal G}}(\omega^{+})|i\rangle$.
${\cal G}_{ii}(\omega^{+})$ for $i\neq c$ may be easily evaluated
using Eq. (\ref{eq:GM-2}), to get 
\[
{\cal{ G}}_{ii}(\omega^{+})=(\omega+i\gamma-\epsilon_{a})^{-1}+V^{2}(\omega+i\gamma-\epsilon_{a})^{-2}{\cal{G}}_{cc}(\omega),
\]
so that 
\begin{equation}
\sum_{i=1}^{N}{\cal G}_{ii}(\omega^{+})=N(\omega+i\gamma-\epsilon_{i})^{-1}+{\cal N}\tilde{V}^{2}{\cal G}_{cc}(\omega^{+})(\omega+i\gamma-\epsilon_{a})^{-2}.\label{eq:TrG}
\end{equation}
A change in the total density of states for the molecules alone may
be defined by

\[
<\Delta\rho_{M}(\omega)>_{\boldsymbol{\xi}} =-\frac{1}{\pi}Im\left(\sum_{i=1}^{N}G_{ii}(\omega^{+})\right)-\rho_{M}^{0}(\omega),
\]
where $\rho_{M}^{0}(\omega)$ is the density of molecular states in
the case where $\tilde{V}=0$. The above may be calculated using Eq.
(\ref{eq:TrG}) to get 
\begin{equation}
\left\langle \Delta\rho_{M}(\omega)\right\rangle _{\boldsymbol{\xi}}=-\frac{1}{\pi}Im\left({\cal N}\tilde{V}^{2}{\cal G}_{cc}(\omega^{+})(\omega+i\gamma-\epsilon_{a})^{-2}\right),\label{Eq.drhom}
\end{equation}
so that the change in the total density of states, due to the interaction
of the cavity mode with the molecules is 
\begin{equation}
\left\langle \Delta\rho_{T}(\omega)\right\rangle _{\boldsymbol{\xi}}=-\frac{1}{\pi}Im\{{\cal G}_{cc}(\omega^{+})\}-\frac{1}{\pi}Im\left({\cal N}\tilde{V}^{2}{\cal G}_{cc}(\omega^{+})(\omega+i\gamma-\epsilon_{a})^{-2}\right).\label{eq:Delta-rho_T}
\end{equation}

\subsection{The Absorption Spectrum}

The Hamiltonian for the interaction of the system with radiation is
\[
\hat{H}_{int}(t)=-\sum_{i=1}^N\boldsymbol{E}_{i}(t).\boldsymbol{\hat{\mu}}_{i}.
\]
The $\boldsymbol{\hat{\mu}}_{i}$ operator can be written in terms
in terms of transition dipole moment as: 
$
\boldsymbol{\hat{\mu}}_{i}=\boldsymbol{\mu_{i}}|g\rangle\langle i|+\boldsymbol{\mu_{i}^{*}}|i\rangle\langle g|
$
where $\boldsymbol{\mu_{i}}=\langle i|\boldsymbol{\hat{\mu}}_{i}|g\rangle$.
We can rewrite the Hamiltonian for interaction as: 
%\begin{equation}
$\hat{H}_{int}(t)=-\sum_{i}\boldsymbol{E}_{i}(t).\left(\boldsymbol{\mu_{i}}|g\rangle\langle i|+\boldsymbol{\mu_{i}^{*}}|i\rangle\langle g|\right).\label{Eq.Hint}$
%\end{equation}
$|g\rangle$ indicates the ground state in which all the molecules
are in their ground states. $|i\rangle$ is the site in which the
$i^{th}$ molecule is excited. $\boldsymbol{E}_{i}(t)$ is the harmonic
electric field of the frequency $\omega$ that the $i^{th}$ molecule
experiences. In the spirit of the model that assumes $V_{i}=V$ a
constant, we first take the electric field to be the same for all
the molecules - i.e., neglect its space dependence so that all the
molecules experience the same electric field. Further, for simplicity,
we take all of them to be oriented in the direction of the electric
field. Under these assumptions,  
\[
\hat{H}_{int}(t)=-E(t)\mu_{eg}\sum_{i}\left(|g\rangle\langle i|+|i\rangle\langle g|\right)
\]
where $\mu_{eg}=|\boldsymbol{\mu_{i}}|$ and $E(t)$ is the electric
field.
The absorption cross section $\alpha(\omega)$ for radiation of frequency
$\omega$ is then 
\[
\alpha(\omega)=\frac{\pi\omega}{\epsilon_{o}c\hbar}\;\;\sum_{f}\left|\left\langle f\left|\hat{H}_{int}(0)\right|g\right\rangle \right|^{2}\delta(E_{f}-E_{g}-\hbar\omega).
\]

In the above, $E_{g}$ is the energy of the ground state. $|f\rangle$
denotes a possible final state, having energy $E_{f}$. Following
reference \cite{schatz2012} we write this as the Fourier transform
of a correlation function: 
\[
\alpha(\omega)=\frac{\omega}{2\epsilon_{o}c\hbar}\int_{-\infty}^{\infty}dte^{i\omega t}\left\langle g\left|\hat{H}_{int}(t)\hat{H}_{int}\right|g\right\rangle ,
\]
where $\hat{H}_{int}(t)=e^{i\hat{H}t/\hbar}\hat{H}_{int}(0)e^{-i\hat{H}t/\hbar}$.
On using the expression for $H_{int}$ and simplifying we get 
\begin{align}
\alpha(\omega) & =\frac{\omega}{2\epsilon_{o}c\hbar}\left|\boldsymbol{{\mu}}\right|^{2}\int_{-\infty}^{\infty}dte^{i\omega t}\sum_{i,j}\left\langle i\left|e^{i\hat{H}t/\hbar}\right|j\right\rangle \nonumber \\
 & =-\frac{\omega}{\epsilon_{o}c\hbar}N\left|\boldsymbol{{\mu}}\right|^{2}\;Im\left[G_{mol,mol}(\omega)\right].\label{eq:Eq.CS}
\end{align}
On performing the average over disorder, we get 

\begin{equation}\left\langle \alpha(\omega)\right\rangle_{\boldsymbol{\xi}}=-\left [ \frac{\omega}{\epsilon_{o}c\hbar}N\left|\boldsymbol{{\mu}}\right|^{2}\;Im\left[{\cal G}_{mol,mol}(\omega)\right] \right ].
\end{equation}
\subsection{Hexagonal arrangement of sites with Cauchy disorder }
\begin{figure}[h]
	\includegraphics[width=0.5\linewidth]{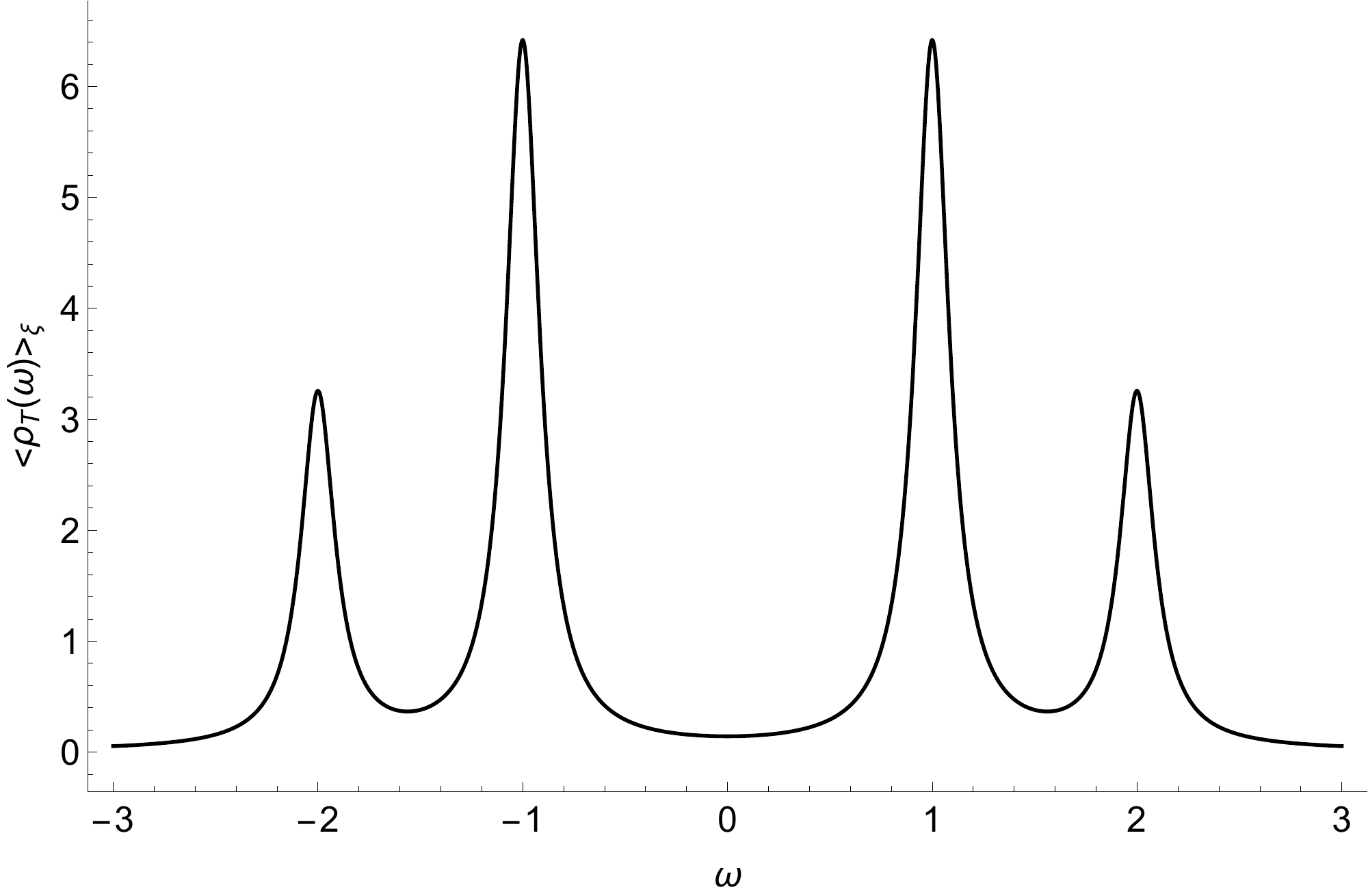} \caption{Plot of $\rho_{T}(\omega)$ against $\omega$ for hexagonal arrangement of sites and $\gamma=0.1$. The two peaks in middle are 2 times more intense than each of the side peaks.  \label{FigrhoHex} }
\end{figure}

Here we give results for a hexagonal arrangement of sites with nearest neighbor hopping, in presence of disorder.  We use the formalism of the paper to get the plot in Fig. \ref{FigrhoHex}.  
For the non-disordered system, the Hamiltonian and the results would be the same as the those of H$\ddot{u}$ckel theory for benzene.  The result of the calculation would be four energy levels, the middle two being doubly degenerate.  In Fig.  we see that the disorder broadens all the levels to the same extent The positions of the peaks are at the locations of the original four levels.  \bibliography{ms}